\documentclass[]{pasj01}

\Received{2024/04/28}
\Accepted{2024/08/11}
 
\usepackage{url}
\usepackage[switch,mathlines]{lineno}

\begin{document} 

\title{Discovery of a hyperluminous quasar at {\boldmath $z$} = 1.62 with Eddington ratio {\boldmath $>$} 3 in the eFEDS field confirmed by KOOLS-IFU on Seimei Telescope}

 \author{
   Yoshiki		\textsc{Toba}			\altaffilmark{1,2,3,4$^{\ast,\dag}$},
   Keito		\textsc{Masu}			\altaffilmark{2},    
   Naomi		\textsc{Ota}			\altaffilmark{2},  
   Zhen-Kai 	\textsc{Gao}			\altaffilmark{3,5},  
   Masatoshi 	\textsc{Imanishi}		\altaffilmark{1},   
   Anri			\textsc{Yanagawa}		\altaffilmark{2},  
   Satoshi 		\textsc{Yamada}			\altaffilmark{6},
   Itsuki		\textsc{Dosaka}			\altaffilmark{7},                
   Takumi		\textsc{Kakimoto}		\altaffilmark{8,1},  
   Seira		\textsc{Kobayashi}		\altaffilmark{7},  
   Neiro		\textsc{Kurokawa}		\altaffilmark{2},  
   Aika			\textsc{Oki}			\altaffilmark{9,10},      
   Sorami		\textsc{Soga}			\altaffilmark{2},    
   Kohei		\textsc{Shibata}		\altaffilmark{7},  
   Sayaka		\textsc{Takeuchi}		\altaffilmark{2},    
   Yukana		\textsc{Tsujita}		\altaffilmark{2},        
   Tohru		\textsc{Nagao}			\altaffilmark{4},
   Masayuki		\textsc{Tanaka}			\altaffilmark{1,8}, 
   Yoshihiro 	\textsc{Ueda}	   		\altaffilmark{11},
   Wei-Hao 		\textsc{Wang}	   		\altaffilmark{3}        
}
\email{yoshiki.toba@nao.ac.jp}
\footnotetext[$^\dag$]{NAOJ fellow}

\altaffiltext{1}{National Astronomical Observatory of Japan, 2-21-1 Osawa, Mitaka, Tokyo 181-8588, Japan}
\altaffiltext{2}{Department of Physics, Nara Women's University, Kitauoyanishi-machi, Nara, Nara 630-8506, Japan}
\altaffiltext{3}{Academia Sinica Institute of Astronomy and Astrophysics, 11F of Astronomy-Mathematics Building, AS/NTU, No.1, Section 4, Roosevelt Road, Taipei 10617, Taiwan}
\altaffiltext{4}{Research Center for Space and Cosmic Evolution, Ehime University, 2-5 Bunkyo-cho, Matsuyama, Ehime 790-8577, Japan}
\altaffiltext{5}{Graduate Institute of Astronomy, National Central University, 300 Zhongda Road, Zhongli, Taoyuan 32001, Taiwan}
\altaffiltext{6}{RIKEN Cluster for Pioneering Research, 2-1 Hirosawa, Wako, Saitama 351-0198, Japan}
\altaffiltext{7}{Department of Physics, Ehime University, 2-5 Bunkyo-cho, Matsuyama, Ehime 790-8577, Japan}
\altaffiltext{8}{Department of Astronomical Science, The Graduate University for Advanced Studies, SOKENDAI, 2-21-1 Osawa, Mitaka, Tokyo 181-8588, Japan}
\altaffiltext{9}{Department of Astronomy, Graduate School of Science, The University of Tokyo, 7-3-1 Hongo, Bunkyo-ku, Tokyo 113-0033, Japan}
\altaffiltext{10}{Mizusawa VLBI Observatory, National Astronomical Observatory Japan, 2-21-1 Osawa, Mitaka, Tokyo 181-8588, Japan}
\altaffiltext{11}{Department of Astronomy, Kyoto University, Kitashirakawa-Oiwake-cho, Sakyo-ku, Kyoto 606-8502, Japan}






\maketitle

\begin{abstract}
We report the discovery of a hyperluminous type 1 quasar (eFEDS J082826.9--013911; eFEDSJ0828--0139) at $z_{\rm spec}$ = 1.622 with a super-Eddington ratio ($\lambda_{\rm Edd}$). 
We perform the optical spectroscopic observations with KOOLS-IFU on the Seimei Telescope.
The black hole mass ($M_{\rm BH}$) based on the single-epoch method with Mg{\,\sc ii} $\lambda$2798 is estimated to be $M_{\rm BH} = (6.2 \pm 1.2) \times 10^8$ $M_{\odot}$.
To measure the precise infrared luminosity ($L_{\rm IR}$), we obtain submillimeter data taken by SCUBA-2 on JCMT and conduct the spectral energy distribution analysis with X-ray to submillimeter data.
We find that $L_{\rm IR}$ of eFEDSJ0828--0139 is $L_{\rm IR} = (6.8 \pm 1.8) \times 10^{13}$ $L_{\odot}$, confirming the existence of a hypeluminous infrared galaxy (HyLIRG).
$\lambda_{\rm Edd}$ is estimated to be $\lambda_{\rm Edd} = 3.6 \pm 0.7$, making it one of the quasars with the highest BH mass accretion rate at cosmic noon.
\end{abstract}

\section{Introduction}
\label{Intro}

It is widely accepted that almost all massive galaxies contain supermassive black holes (SMBHs) with a BH mass ($M_{\rm BH}$) of 10$^{5-10}$ $M_{\odot}$ at the galaxy center and that their masses are strongly correlated with the mass of the spheroidal component of galaxies (e.g., \cite{Magorrian,Ferrarese,Marconi,Kormendy,McConnell}).
This suggests that the formation of SMBHs is closely related to the formation of galaxies, manifesting their co-evolution with host galaxies.
However, the understanding of the co-evolution mechanism between the two, which differ by 9--10 orders of magnitude on the physical scale, has not been observationally constrained.

To address this issue, we focus on hyperluminous infrared galaxies (HyLIRGs) with an infrared (IR) luminosity ($L_{\rm IR}$) of $> 10^{13} L_{\odot}$ \citep{Rowan}, most IR emission comes from hot dust heated by active galactic nuclei (AGNs) (e.g., \cite{Toba15,Symeonidis}).
According to the galaxy and SMBH growth scenarios resulting from galaxy mergers expected from numerical simulations, HyLIRGs ``theoretically'' correspond to the most interesting phases in which the growth rates of galaxies and SMBH peak (e.g., \cite{Narayanan,Blecha,Yutani}).
Therefore, HyLIRGs are expected to serve as a crucial laboratory for probing the growth phase of co-evolution.
However, AGN and host properties in HyLIRGs have still been poorly understood ``observationally'' because the number density of HyLIRGs is very small, and there have not been extensive multi-wavelength observations for them.
To overcome this situation, deep and wide-field observation, particularly with X-ray data, is invaluable for uncovering the AGN properties of a spatially rare population, HyLIRGs, although X-ray observations of HyLIRGs are still limited (e.g., \cite{Wilman,Hlavacek,Ricci,Toba20a,Toba21b}).

\begin{table}[t]
\tbl{Observed Properties of eFEDSJ0828$-$0139.}{
\scalebox{0.95}[1.0]{
\begin{tabular}{lr}
\hline
\hline
eFEDS J082826.9$-$013911				&										\\
\hline
ID\_SRC	\citep{Brunner}					&	14050								\\
SDSS ObjID								&	1237655176541241479					\\
R.A. (LS8) [J2000.0] 					&	08:28:27.145						\\
Decl. (LS8) [J2000.0]					&	$-$01:39:12.18  					\\
Redshift ($z_{\rm spec}$)				&	1.6224 $\pm$ 0.0007					\\
\hline
eROSITA $f_{\rm 2-10\,keV}$  [mJy]		&	$(8.83 \pm 3.64) \times 10^{-7}$	\\
eROSITA $f_{\rm 0.5-2\,keV}$ [mJy]		&	$(4.06 \pm 1.67) \times 10^{-6}$	\\
GALEX FUV [mJy]							&	$(1.13 \pm 0.45) \times 10^{-2}$	\\
GALEX NUV [mJy]							&	$(7.13 \pm 0.73) \times 10^{-2}$	\\
SDSS $u$-band [mJy]						&	$(2.54 \pm 0.04) \times 10^{-1}$	\\
SDSS $g$-band [mJy]						&	$(5.32 \pm 0.02) \times 10^{-1}$	\\
SDSS $r$-band [mJy]						&	$(7.64 \pm 0.03) \times 10^{-1}$	\\
SDSS $i$-band [mJy]						&	$(8.84 \pm 0.04) \times 10^{-1}$	\\
SDSS $z$-band [mJy]						&	$(8.45 \pm 0.09) \times 10^{-1}$	\\
UKIDSS $Y$-band [mJy]					&	$(7.26 \pm 0.10) \times 10^{-1}$	\\
UKIDSS $J$-band [mJy]					&	$(8.31 \pm 0.07) \times 10^{-1}$	\\
UKIDSS $H$-band [mJy]					&	$(10.4 \pm 0.12) \times 10^{-1}$	\\
UKIDSS $K$-band [mJy]					&	$(7.60 \pm 0.13) \times 10^{-1}$	\\
WISE 3.4 $\mu$m [mJy]					&	$(11.4 \pm 0.05) \times 10^{-1}$	\\
WISE 4.6 $\mu$m [mJy]					&	2.88 $\pm$ 0.12						\\
WISE 12  $\mu$m [mJy]					&	3.30 $\pm$ 0.16						\\
WISE 22  $\mu$m [mJy]					&	11.2 $\pm$ 1.29						\\
AKARI 65 $\mu$m [Jy]					&	$< 1.44$\footnotemark[$*$]			\\
AKARI 90 $\mu$m [Jy]					&	$< 0.33$\footnotemark[$*$]			\\
AKARI 140 $\mu$m [Jy]					&	$< 0.84$\footnotemark[$*$]			\\
AKARI 160 $\mu$m [Jy]					&	$< 3.78$\footnotemark[$*$]			\\
SCUBA-2/JCMT 450 $\mu$m [mJy]			&	$< 44.6$\footnotemark[$*$]			\\
SCUBA-2/JCMT 850 $\mu$m [mJy]			&	$< 5.90$\footnotemark[$*$]			\\
\hline
$E(B-V)_{*}$							&	(1.3 $\pm$ 0.4) $\times 10^{-1}$	\\
$M_*$ [$M_{\odot}$]						&	(3.9 $\pm$ 2.0) $\times 10^{11}$	\\
SFR [$M_{\odot}$ yr$^{-1}$]				&	(1.3 $\pm$ 0.5) $\times 10^3$ 		\\
$L_{\rm IR}$ [$L_\odot$]				&	(6.8 $\pm$ 1.8) $\times 10^{13}$	\\
\hline
$L^{\rm AGN}_{\rm bol}$ [erg s$^{-1}$]	&	(2.9 $\pm$ 0.1) $\times10^{47}$		\\
$M_{\rm BH}$ [$M_{\odot}$]				& 	(6.2 $\pm$ 1.2) $\times 10^8$		\\ 
$\lambda_{\rm Edd}$						&	$3.6 \pm 0.7$						\\
\hline
\end{tabular}
}
}
\label{sample}
\begin{tabnote}
The SDSS ObjID is valid for SDSS Data Release 8 or later. All the flux densities from X-ray (eROSITA) to mid-IR (WISE) are corrected for Galactic extinction. \\
\footnotemark[$*$]3$\sigma$ upper limits.
\end{tabnote}
\end{table}

The advent of the eROSITA X-ray satellite \citep{Merloni12,Predehl} has enabled us to systematically investigate HyLIRGs from an AGN point of view because X-ray radiation is sensitive to finding AGNs.
In this paper, we report the discovery of a hyperluminous quasar, eFEDS J082826.9$-$013911 (hereafter, eFEDSJ0828--0139), at $z_{\rm spec}$ = 1.622 with a supper-Eddington ratio of $\lambda_{\rm Edd} > 3.0$.
Table \ref{sample} summarizes complete information, including photometry on this object and its physical properties obtained in this work.
This is confirmed by (i) optical spectroscopic observations with the Kyoto Okayama Optical Low-dispersion Spectrograph with optical fiber IFU (KOOLS-IFU: \cite{Yoshida,Matsubayashi}) on the Seimei Telescope \citep{Kurita}, and (ii) multi-wavelength data analysis from X-ray to submillimeter.
In particular, because IFU data mitigate slit-loss of flux, unlike single-fiber or -slit spectroscopy, our KOOLS-IFU observation benefits not just extended sources like nearby galaxies (e.g., \cite{Toba22a}) but also point sources like AGNs  (e.g., \cite{Hoshi}).

The remainder of this paper is structured as follows.
We describe target selection and follow-up observations using optical spectroscopy and submillimeter imaging in section \ref{DA}.
The derived AGN and host properties with their potential uncertainties and the characterization of eFEDSJ0828--0139 are presented in section \ref{RD}.
We summarize the results of this work in section \ref{Sum}.
This work assumes a flat Universe with $H_0$ = 70 km s$^{-1}$ Mpc$^{-1}$, $\Omega_M$ = 0.3, and $\Omega_\Lambda$ = 0.7.

\section{Data and analysis}
\label{DA}

\subsection{Sample selection}

The target (eFEDSJ0828--0139) was selected from the HyLIRG candidate sample provided by \citet{Toba22b}.
A full description of HyLIRG candidate selection is presented by Y. Toba et al. (in preparation).
Hence, we present a summary.
\citet{Toba22b} performed the spectral energy distribution (SED) analysis for Wide-field Infrared Survey Explorer (WISE: \cite{Wright}) 22 $\mu$m-selected sources in the eROSITA Final Equatorial Depth Survey (eFEDS: \cite{Brunner}),
According to their $L_{\rm IR}$, 150 HyLIRG candidates were left, more than half of which have spectroscopic redshift ($z_{\rm spec}$) through the follow-up campaign, such as the Sloan Digital Sky Survey (SDSS; \cite{York}) V/eFRDS observations \citep{Almeida}.

Our target is the brightest source in the optical ($r$-mag = 16.78) among our HyLIRG candidate sample with photometric redshift ($z_{\rm phot}$).
eFEDSJ0828--0139 is classified as an unobscured AGN (i.e., quasar) with a hydrogen column density of $\log\,N_{\rm H} \sim 20.8$ cm$^{-2}$ through an eROSITA spectral analysis \citep{Liu}.
This object has also been selected as an AGN/quasar candidate from a multi-wavelength photometric perspective, such as WISE \citep{Secrest,Assef} and Gaia \citep{Bailer-Jones,Wu23}.
Its $z_{\rm phot}$ is estimated by some studies, for instance, $z_{\rm phot} = 2.825^{+0.405}_{-0.215}$ \citep{Richards}, $1.321 \pm 0.611$ \citep{Duncan} and $1.367^{+0.192}_{-0.126}$ \citep{Salvato}, indicating a large redshift uncertainty.

\subsection{Observations and data reduction}

\subsubsection{KOOLS-IFU on Seimei telescope}

To measure $z_{\rm spec}$ and BH properties, we observed eFEDSJ0828--0139 with the KOOLS-IFU on the Seimei Telescope in 2023 (PI: Y.Toba with proposal IDs = 23A-N-CN01 and 23B-N-CN07).
The Seimei Telescope is a 3.8-meter diameter optical-IR alt-azimuth mount telescope located at Okayama Observatory, Kyoto University, Okayama Prefecture in Japan, where the typical seeing on this site is 1.2\arcsec--1.4\arcsec.
The KOOLS-IFU comprises 117 fibers with a total field of view (FoV) of 8.4\arcsec $\times 8.0\arcsec$\footnote{The configuration of KOOLS-IFU (such as number of fibers) was updated in October 2020.}.
We used the VPH-blue\footnote{\url{http://www.o.kwasan.kyoto-u.ac.jp/inst/p-kools/performance/}} among four grisms equipped with KOOLS-IFU.
The wavelength coverage is 4100--8900\,\AA\, and the spectral resolution ($R = \lambda/\Delta \lambda$) is approximately 500.
The observational log is summarized in Table \ref{log}.
The total integration time is approximately 4.3 hours.

\begin{table}[t]
\tbl{Observation log in 2023A and 2023B semester.}{
\scalebox{0.85}[0.9]{
\begin{tabular}{lccl}
\hline\hline
\multicolumn{1}{c}{Observing date} &	Grism	& \multicolumn{1}{c}{Exposure time (s)}  & \multicolumn{1}{c}{Standard star} \\
\hline
Jan. 25, 2023 & VPH-blue & 7200 & HR1544 \\
Jan. 27, 2023 & VPH-blue & 3600 & HD74280 \\
Feb. 13, 2023 & VPH-blue &  600 & HD74280 \\
Dec. 12, 2023 & VPH-blue & 4200 & HD74280 \\ 
\hline
\end{tabular}
}
}
\label{log}
\end{table}

We note that our target is well-fitted by a round exponential galaxy model (REX\footnote{\url{https://www.legacysurvey.org/dr8/catalogs/#goodness-of-fits-and-morphological-type}}) rather than a point spread function (PSF) model according to the DESI Legacy Imaging Survey catalog \citep{Dey}, which could indicate that the target is a slightly extended source.
The Petrosian \citep{Petrosian} radius in $r$-band is 1.99 $\pm$ 0.01 arcsec (see {\tt PhotoObj} table in the SDSS).
Hence, our target would benefit from an IFU observation to mitigate flux loss.

The data reduction was executed with the Image Reduction and Analysis Facility (IRAF: \cite{Tody}) and the pipeline tools\footnote{We downloaded a package on April 11, 2024 (see \url{http://www.kusastro.kyoto-u.ac.jp/~iwamuro/KOOLS/}).} dedicated to the KOOLS-IFU.
This procedure includes overscan subtraction, bad column correction, bias subtraction, flat fielding, wavelength calibration, spectral extraction, sky subtraction, flux calibration, and making a spectrum by combining spectra from all fibers on which the object is located.
Ne, Hg, and Xe lamp data were used for wavelength calibration.
We took a weighted mean of the spectra taken each day to obtain a spectrum with a high signal-to-noise (SN) ratio.
Possible flux loss was corrected by scaling the spectrum to match the SDSS $r$-band magnitude.

\subsubsection{SCUBA-2 on JCMT}
\label{ss_SCUBA2}

In \citet{Toba22b}, the far-IR (FIR) and submillimeter data for eFEDSJ0828--0139 were lacking, and only AKARI FIR (shallow) upper limits were provided.
Because the submillimeter data are crucial for precise measurement of $L_{\rm IR}$ (\cite{Toba18,Toba20b}), we observed eFEDSJ0828$-$0139 with the Submillimetre Common User Bolometer Array 2 (SCUBA-2: \cite{Holland}) on the James Clerk
Maxwell Telescope (JCMT), providing 450 and 850 $\mu$m photometry.
Our observation was executed under the Band-1 condition ($\tau_{\rm 225\,GHz} < 0.05$) on January 23, 2024, through a 24A regular program (M24AP001, PI: Y.Toba). 
The total on-source integration time is about 1 hour, twice the 30-minute scans by the compact ``Daisy'' scan pattern. 
During the observations, we observed a nearby radio source, 0828+046, for a pointing check. 
Data reduction and flux measurements at 450 and 850 $\mu$m were performed in a standard manner (e.g., \cite{Wang,Lim,Zhen-Kai}) with the aid of the Sub-Millimeter Common User Reduction Facility \citep{Chapin} and the Pipeline for Combining and Analyzing Reduced Data (PICARD: \cite{Jenness}). 
A full description of SCUBA-2 data reduction and flux measurements will be provided in Y.Toba et al. (in preparation)\footnote{Our observations aim at observing seven HyLIRG candidates, including eFEDSJ0828$-$0139, which will be presented once observations are completed.}.

\subsection{Spectral fitting}

To derive AGN properties such as AGN bolometric luminosity ($L^{\rm AGN}_{\rm bol}$), $M_{\rm BH}$, and $\lambda_{\rm Edd}$ of our quasar, spectral fitting to KOOLS-IFU spectrum was conducted using the Quasar Spectral Fitting package ({\tt QSFit} v1.3.0\footnote{\url{https://qsfit.inaf.it}}: \cite{Calderone}).
We fitted the KOOLS-IFU spectrum as a combination of (i) an AGN continuum with a single power-law, (ii) a Balmer continuum modeled by \citet{Grandi} and \citet{Dietrich}, (iii) iron-blended emission lines with UV-optical templates \citep{Veron,Vestergaard01}, and (iv) emission lines with Gaussian components.
{\tt QSFit} fits all the components simultaneously following a Levenberg-Marquardt least-squares minimization algorithm with {\tt MPFIT} \citep{Markwardt} procedure.
Spectral fitting was executed after correcting for the galactic extinction provided by \citet{SFD} with the Milky Way attenuation curve \citep{O'Donnell}.

The main purpose of this spectral fitting is to measure the full width at half maximum (FWHM) of Mg{\,\sc ii}\,$\lambda$2798  and continuum luminosity at 3000\,\AA, $\lambda\,L_{\lambda} {\rm (3000\,\AA)}$, that are ingredients for $M_{\rm BH}$ estimates. 
Based on outputs from {\tt QSFit}, we estimated $M_{\rm BH}$ through a single-epoch recipe provided by \citet{Vestergaard};
\begin{equation}
\label{Eq}
M_{\rm BH}\,[M_{\odot}] =   10^{6.86} \left[\frac{\rm FWHM\,(MgII)}{1000 \,{\rm km \, s^{-1}}} \right]^2 \left[\frac{\lambda L_{\lambda}\, ({\rm 3000\, \AA})}{10^{44} \,{\rm erg \, s^{-1}}}\right]^{0.5}.
\end{equation}
$L^{\rm AGN}_{\rm bol}$ was converted from BC$_{\rm 3000} \times \lambda L_{\lambda}\,{\rm (3000\,\AA)}$ where BC$_{\rm 3000}$ = $5.2 \pm  0.2$ is the bolometric correction \citep{Runnoe}.
Following \citet{Toba21a}, the uncertainty in $M_{\rm BH}$ is calculated through the error propagation of Equation \ref{Eq} while the uncertainty in $L^{\rm AGN}_{\rm bol}$ is propagated from 1$\sigma$ errors of $\lambda L_{\lambda}\,{\rm (3000\,\AA)}$ and BC$_{\rm 3000}$.

\subsection{SED fitting}

We collected the multi-wavelength data from X-ray to FIR in the same manner as \citet{Toba22b}.
We refer the reader to that paper for details, but in short, we used GALEX \citep{Martin} for ultraviolet (UV) data, SDSS for optical data, UKIDSS \citep{Lawrence} for near-IR (NIR) data, WISE for mid-IR (MIR) data, and AKARI \citep{Murakami} for FIR data.
UV-to-MIR photometry is corrected for Galactic extinction \citep{SFD}.
In addition, we have added photometry obtained from SCUBA-2 observation (section \ref{ss_SCUBA2}).
Because our target was not detected at 450 and 850 $\mu$m, we input their 3$\sigma$ upper limits for the SED fitting, which is crucial to pin down the FIR SED of eFEDSJ0828$-$0139.

We employed the Code Investigating GALaxy Emission ({\tt CIGALE} ver.2022.1\footnote{\url{https://cigale.lam.fr/2022/07/04/version-2022-1/}}; \cite{Burgarella,Noll,Boquien,Yang,Yang22}).
This code allows us to include values for many parameters related to, e.g., the star formation (SF) history (SFH), single stellar population (SSP), attenuation law, AGN emissions, and dust emissions by considering the energy balance between the UV/optical and IR (see, e.g., \cite{Hashiguchi,Toba24}). 
Table \ref{Param} summarizes the parameter ranges used in the SED fitting with {\tt CIGALE}.
These parameter sets are the same as presented in \citet{Toba21b} optimized for HyLIRG candidates.
A full description of the assumed models is provided by \citet{Toba21b}; hence, we provide a summary.
We assume a delayed SFH with recent starburst \citep{Ciesla} with parameterizing e-folding time of the main stellar population model ($\tau_{\rm main}$), the age of the main stellar population in the galaxy, the age of the burst, and the ratio of the SF rate (SFR) after and before the burst (R\_sfr).
A starburst attenuation curve \citep{Calzetti,Leitherer} is used for dust attenuation, in which we parameterize the color excess of the nebular emission lines, $E(B-V)_{\rm lines}$.
We chose the SSP model \citep{Bruzual}, assuming the initial mass function (IMF) of \cite{Chabrier}, and the standard nebular emission model with the implementation of the new CLOUDY HII-region model \citep{Villa} is included in {\tt CIGALE} \citep{Inoue}.
AGN emission is modeled by using the {\tt SKIRTOR} \citep{Stalevski}.
This torus model consists of seven parameters: torus optical depth at 9.7 $\mu$m ($\tau_{\rm 9.7}$), torus density radial parameter ($p$), torus density angular parameter ($q$), the angle between the equatorial plane and edge of the torus ($\Delta$), the ratio of the maximum to minimum radii of the torus ($R_{\rm max}/R_{\rm min}$), viewing angle ($\theta$), and AGN fraction in total IR luminosity ($f_{\rm AGN}$).
Dust grain emission is modeled by \citet{Draine} in which we parameterize the mass fraction of PAHs ($q_{\rm PAH}$), the minimum radiation field ($U_{\rm min}$), the power-low slope of the radiation field distribution ($\alpha$), and the fraction illuminated with a variable radiation field ranging from $U_{\rm min}$ to $U_{\rm max}$ ($\gamma$).
X-ray emission is modeled with a fixed power-law photon index of AGN \citep{Liu}, and only $\alpha_{\rm OX}$ is parameterized.

\begin{table}
\tbl{Parameter values used in SED fitting with {\tt CIGALE}}{
\scalebox{0.95}[1.0]{
\begin{tabular}{lc}
\hline \hline
Parameter & Value\\
\hline
\multicolumn{2}{c}{Delayed SFH with recent starburst \citep{Ciesla}}\\
\hline
$\tau_{\rm main}$ [Gyr] & 1.0, 4.0, 8.0, 12 \\
age [Gyr] 				& 0.5, 1.0, 1.5, 2.0 \\
age of burst [Myr] 		& 10, 50, 100 \\
R\_sfr					& 1, 5, 10 \\
\hline
\multicolumn{2}{c}{SSP \citep{Bruzual}}\\
\hline
IMF				&	\cite{Chabrier} \\
Metallicity		&	0.02  \\
\hline
\multicolumn2c{Nebular emission \citep{Inoue}}\\
\hline
$\log\, U$					&	$-$3.0, $-$2.0, $-$1.0	\\
\hline
\multicolumn2c{Dust attenuation \citep{Calzetti,Leitherer}}\\
\hline
$E(B-V)_{\rm lines}$ &  0.0, 0.1, 0.2, 0.3, 0.4, 0.5, 1.0 \\
\hline
\multicolumn{2}{c}{AGN Emission \citep{Stalevski12,Stalevski}}\\
\hline
$\tau_{\rm 9.7}$ 			&  	3, 7, 11					\\
$p$							&	0.5,  1.5					\\
$q$							&	0.5, 1.5					\\
$\Delta$ [$\degree$]		&	40							\\
$R_{\rm max}/R_{\rm min}$ 	& 	30 							\\
$\theta$ [$\degree$]		&	0, 10, 20					\\
$f_{\rm AGN}$ 				& 	0.4, 0.5, 0.6, 0.7, 0.8, 0.9 \\
\hline
\multicolumn{2}{c}{Dust Emission \citep{Draine}}\\
\hline
 $q_{\rm PAH}$ &  2.50, 5.26, 6.63, 7.32 \\
 $U_{\rm min}$ & 10.0, 50.0 \\
 $\alpha$ & 1.0, 1.5, 2.0 \\
 $\gamma$ & 0.01, 0.1, 1.0 \\
\hline
\multicolumn{2}{c}{X-ray Emission \citep{Yang22}}\\
\hline
AGN photon index ($\Gamma$)	&	2.0		\\
$\alpha_{\rm OX}$	&	$-$2.0, $-$1.9, $-$1.8, $-$1.7 \\
$|\Delta\,\alpha_{\rm OX}|_{\rm max}$	&	0.5	\\
\hline
\end{tabular}
}
}
\label{Param}
\end{table}

\section{Results and discussion}
\label{RD}

\subsection{Results of spectral fitting and AGN properties}
\label{dis_spec}

\begin{figure*}
 \begin{center}
  \includegraphics[width=\textwidth]{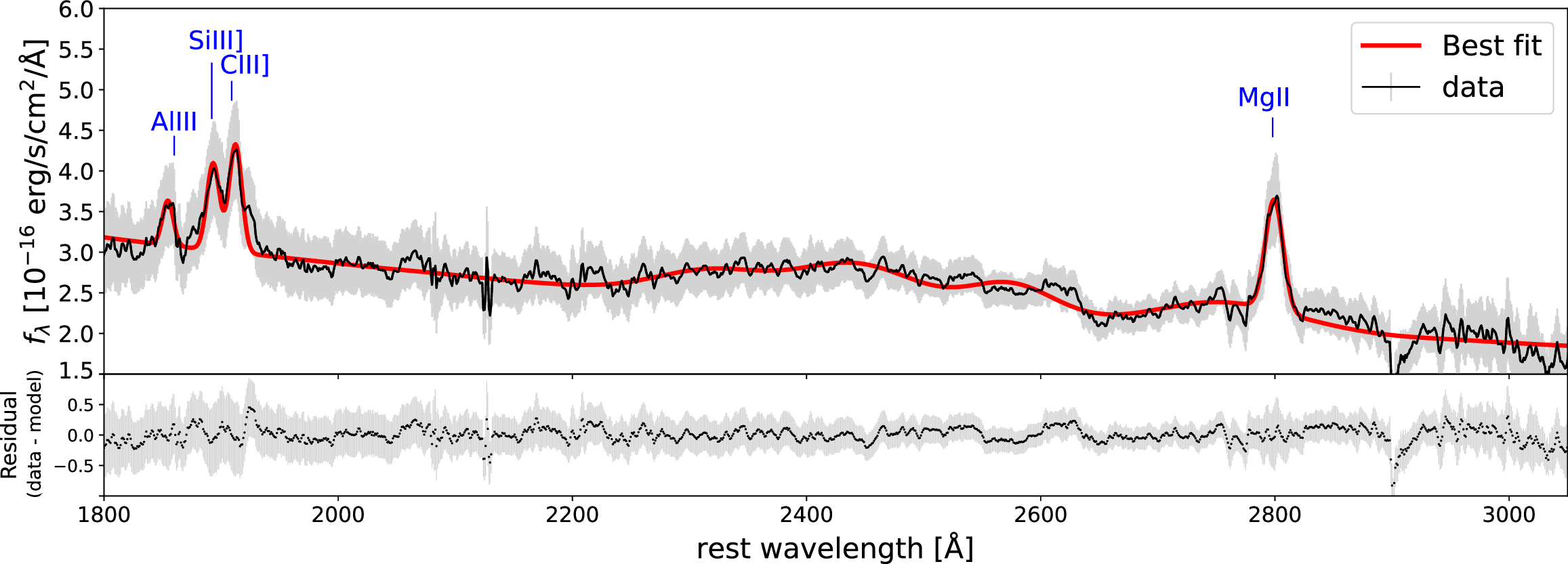} 
 \end{center}
\caption{The optical spectrum of eFEDSJ0828$-$0139 taken by the KOOLS-IFU. The black and gray lines represent observed data and its 1$\sigma$ uncertainty. Blue vertical lines and labels mark detected lines. The best-fit line by {\tt QSFit} is shown with the red line. The bottom panel shows the residual (data -- model) with 1$\sigma$ uncertainty. Note that there is a relatively large residual around rest-frame $\sim$ 2900 \AA\, (observed-frame 7500 \AA). This is due to the strongest telluric absorption line by O$_2$ (A-band) in the observed wavelength \citep{Rudolf}.}
\label{spectrum}
\end{figure*}

Figure \ref{spectrum} shows the optical spectrum of eFEDSJ0828$-$0139 taken by the KOOLS-IFU, where several emission lines such as Al{\,\sc iii}$\lambda$1860, Si{\,\sc iii}]$\lambda$1892, C{\,\sc iii}]$\lambda$1909 and Mg{\,\sc ii} are clearly detected.
The result of the spectral fitting with {\tt QSFit} is also shown in Figure \ref{spectrum}.
The spectroscopic redshift of eFEDSJ0828$-$0139 measured from Mg{\,\sc ii} is $z_{\rm spec} = 1.6224 \pm 0.0007$.
The FWHM of Mg{\,\sc ii} and the continuum luminosity at 3000 \AA\, are FWHM (Mg{\,\sc ii}) = $(1.9 \pm 0.2)
\times 10^3$ km s$^{-1}$ and $\lambda L_{\lambda}\,{\rm (3000\,\AA)} = (56.1 \pm 0.4) \times 10^{45}$ erg s$^{-1}$, respectively, which results in the black hole mass of $M_{\rm BH} = (6.2 \pm 1.2) \times 10^8$ $M_{\odot}$.
We note that \citet{Buendia} recently provided a recipe for virial BH mass based on Al{\,\sc iii} line.
We find that the BH mass using their recipe is $\log\,(M^{\rm AlIII}_{\rm BH}/M_{\odot}) \sim 8.5$, which is roughly consistent with Mg{\,\sc ii}-based $M_{\rm BH}$.
Eddington luminosity is $L_{\rm Edd} = (8.0 \pm 0.2) \times 10^{46}$ erg s$^{-1}$.
The AGN bolometric luminosity is $L^{\rm AGN}_{\rm bol} = (2.9 \pm 0.1) \times 10^{47}$ erg s$^{-1}$, indicating a hyper-luminous quasar as similar to WISE-SDSS selected, WISSH quasars at $z > 2$ \citep{Duras_WISSH}.
Consequently, the Eddington ratio of eFEDSJ0828$-$0139 is estimated to be $\lambda_{\rm Edd} = 3.6 \pm 0.7$, making it a prominent quasar with BH at super-Eddington accretion.

\citet{Mart} reported that such quasars with extremely-high $\lambda_{\rm Edd}$ (so-called extreme accretor (xA) quasars: \cite{Marziani}) shows relatively strong Al{\,\sc iii} and Si{\,\sc iii}] emissions compared with C{\,\sc iii}] and conspicuous  Fe{\,\sc ii} lines in their rest UV-to-optical spectra.
We find that the line flux ratio for Al{\,\sc iii}/Si{\,\sc iii}] and C{\,\sc iii}]/Si{\,\sc iii}] is $0.44 \pm 0.18$ and $1.25 \pm 0.27$, respectively.
These values satisfy the selection criteria of xA quasars \citep{Marziani}, supporting that eFEDSJ0828$-$0139 is an extremely high $\lambda_{\rm Edd}$ quasar.

A caveat is that BH mass estimated by a single-epoch method (i.e., Equation \ref{Eq}) has a sizeable systematic uncertainty up to 0.4 dex (see, e.g., \cite{Shen}), which has also been supported by the recent reverberation mapping with the SDSS \citep{Shen24}.
This means that the estimated Eddington ratio also potentially has a large uncertainty.
We confirm the estimated BH mass is consistent with that using another emission line (albeit a single epoch) and obtain evidence of a high Eddington ratio inferred from emission line ratios.
However, our $M_{\rm BH}$ and $\lambda_{\rm Edd}$ do not include the systematic errors mentioned above, which should be kept in mind for the following discussion.

\subsection{Results of SED fitting and AGN host properties}
\label{R_SED}

\begin{figure}[h]
 \begin{center}
  \includegraphics[width=0.48\textwidth]{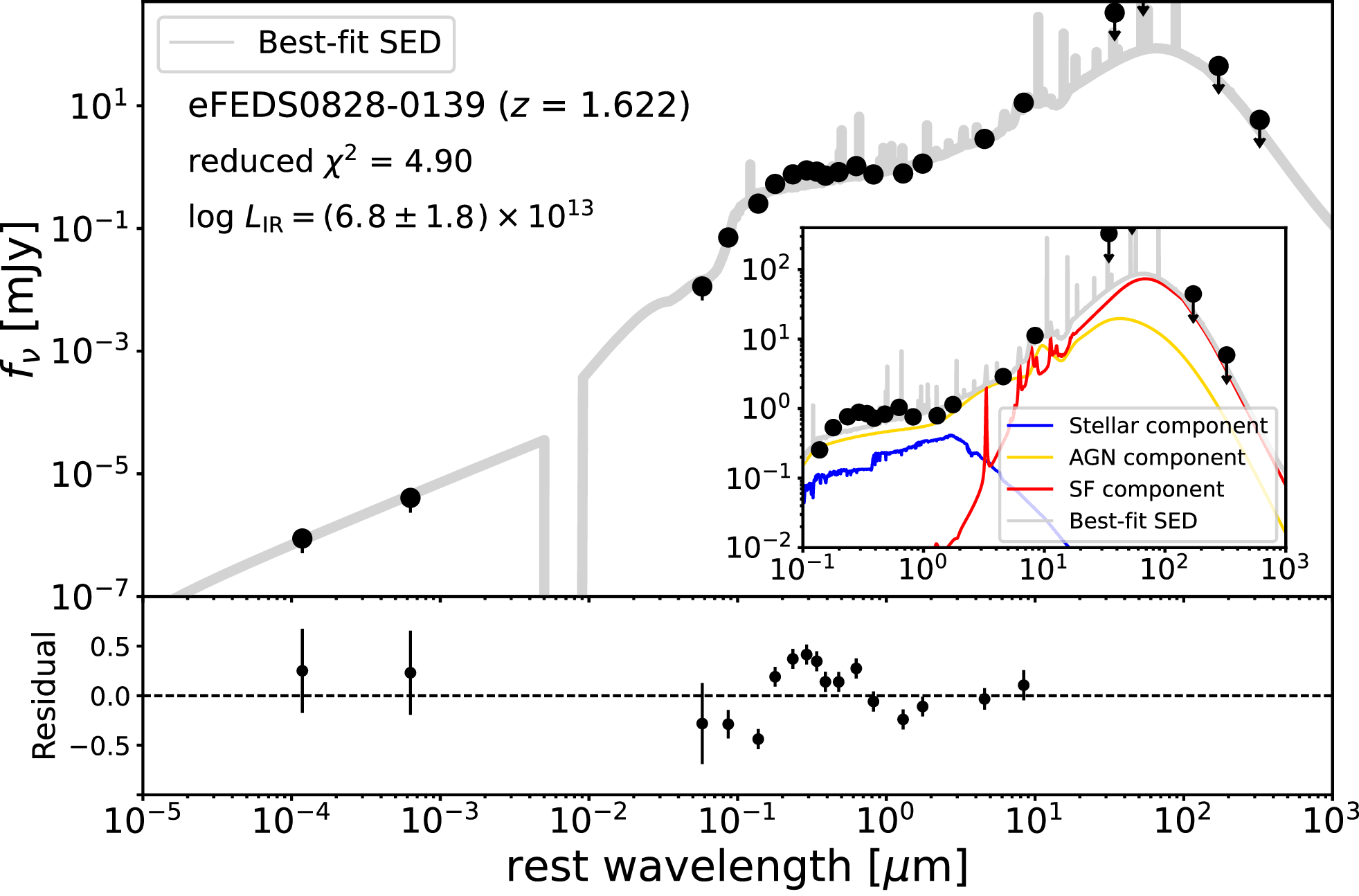} 
 \end{center}
\caption{Best-fit SED of eFEDSJ0828$-$0139. The black points are the photometric data, and the solid gray line represents the resultant best-fit SED. The inset figure shows the SED at 0.1--1000 $\mu$m, where the contributions from the stellar, nebular, AGN, and SF components to the total SED are shown as blue, green, yellow, and red lines, respectively. Relative residual
(defined as (data -- best-fit)/data) are shown at the bottom, where the black line represents the case in which the residual is zero.}
\label{SED}
\end{figure}

Figure \ref{SED} shows the best-fit SED of eFEDSJ0828$-$0139, demonstrating that the observed SED is moderately well-fitted by the stellar, nebular, AGN, and SF components with a reduced $\chi^2$ of 4.9.
The derived total IR luminosity is $L_{\rm IR} = (6.8 \pm 1.8) \times 10^{13}$ $L_{\odot}$, which confirms that our target is an HyLIRG.
The stellar mass ($M_{*}$) and star formation rate (SFR) are $M_* = (3.9 \pm 2.0) \times 10^{11}$ $M_{\odot}$ and SFR = $(1.3 \pm 0.5) \times 10^3$ $M_{\odot}$ yr$^{-1}$, respectively.
These are typical values of HyLIRGs reported in previous works (e.g., \cite{Gao}).
The BH mass and stellar mass ratio of eFEDSJ0828$-$0139 is $\log (M_{\rm BH}/M_*)$ = $-$2.8 $\pm$ 0.2, which agrees with the results reported in \citet{Suh}.

\subsection{Potential uncertainties on IR luminosity and SFR}
\label{Po}

A potential issue caused by the SED fitting with upper limits of AKARI and SCUBA-2 data is about SF (i.e., dust emission from host galaxy) contribution to the total SED, which is also relevant to the accuracy of IR luminosity and SFR.
The {\tt CIGALE} takes into account energy conservation of the amount of UV/optical radiation from SF and AGN absorbed by a dust and the amount of IR radiation re-emitted by the dust when the SED fitting.
Hence, FIR SED is expected to be constrained reasonably.
Nevertheless, the lack of deep FIR data around the peak of the FIR SED would affect the results.
To test the requirement to add an SF component to the SED fitting, we employ the Bayesian information criterion (BIC; \cite{Schwarz}) for two fits that are derived with and without an SF component.
The BIC is defined as BIC = $\chi^{2}$ + $k$ $\times$ ln($n$), where $\chi^{2}$ is non-reduced chi-square, $k$ is the number of degrees of freedom (DOF), and $n$ is the number of photometric data points used for the fitting, respectively.
We then compare the results of two SED fittings without/with the SF (dust emission) module by using $\Delta$BIC = BIC$_{\rm woSF}$ -- BIC$_{\rm wSF}$.
The $\Delta$BIC tells whether the SF/dust model is needed to provide a better fit by considering the difference in DOF (e.g., \cite{Ciesla,Buat,Aufort,Toba20c}).
If $\Delta$BIC is larger than two, adding the SF/dust component provides a better fit than not \citep{Liddle,Stanley}.
The resultant value for eFEDSJ0828$-$0139 is $\Delta$BIC = 7.3, which suggests that the SF/dust component is required to explain the observed SED.

We also estimate IR luminosity and SFR based on the SED fitting without using SCUBA-2 data to see how even the upper limits of SCUBA-2 data are crucial to constrain those quantities.
The resultant values are $L_{\rm IR} = (7.3 \pm 1.9) \times 10^{13}$ $L_{\odot}$ and SFR = $(1.6 \pm 0.6) \times 10^3$ $M_{\odot}$ yr$^{-1}$, which suggests that SCUBA-2 data prevents SFR and $L_{\rm IR}$ from being overestimated.
In summary, dust emission from the host galaxy requires explaining the observed SED, and SCUBA-2 data are crucial to pin down the FIR SED, even if they are upper limits.
Hence, potential uncertainties on IR luminosity and SFR are expected to be small in this work.

\subsection{Characterization of eFEDSJ0828$-$0139}
\label{Dis}

\begin{figure}[h]
 \begin{center}
  \includegraphics[width=0.48\textwidth]{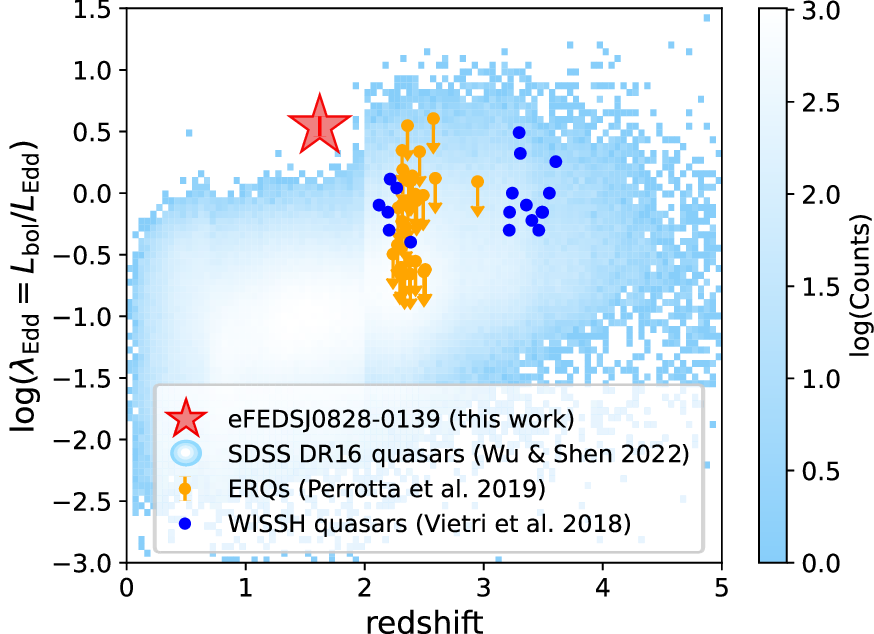} 
 \end{center}
\caption{Eddington ratio as a function of redshift. The red star represents eFEDSJ0828$-$0139. The blue 2D histogram represents
the number density of SDSS DR16 quasars \citep{Wu}. Quasars with $\lambda_{\rm Edd}/\sigma_{\lambda_{\rm Edd}} > 5$ are plotted. The blue and orange circles represent ERQs \citep{Perrotta} and WISSH quasars \citep{Vietri}, respectively.}
\label{Edd}
\end{figure}

Figure \ref{Edd} shows the Eddington ratio as a function of redshift.
We compare the Eddington ratio from a value-added catalog\footnote{\url{http://quasar.astro.illinois.edu/paper_data/DR16Q/dr16q_prop_May16_2023.fits.gz}} \citep{Wu} for the SDSS DR16 quasar catalog \citep{Lyke}, in which the continuum and emission-line properties for 750,414 broad-line quasars are provided.
Note that 25 sources have $\lambda_{\rm Edd} > 3$ among the quasars with $1 < z < 2$.
We visually check their spectra and find that the Eddington ratios of the majority of these quasars are poorly constrained with a large uncertainty of $\lambda_{\rm Edd}$ ($\sigma_{\lambda_{\rm Edd}}$) partially due to the low SN of the emission lines (C{\,\sc iv} and  Mg{\,\sc ii}).
If we restrict ourselves to $\lambda_{\rm Edd}/\sigma_{\lambda_{\rm Edd}} > 5$ (that is similar to eFEDSJ0828$-$0139), only four objects remain.
We also compare $\lambda_{\rm Edd}$ of WISSH quasars \citep{Vietri} and extremely red quasars (ERQs: \cite{Perrotta}) that are also known as high $\lambda_{\rm Edd}$ quasars.

We find that eFEDSJ0828$-$0139 has the highest $\lambda_{\rm Edd}$ at $z \sim$ 1.6, which is even higher than the ERQs and WISSH quasars at $z > 2$.
In addition to a high BH accretion rate, this object has an extremely high SFR ($> 1000$ $M_{\odot}$ yr$^{-1}$) as described in section \ref{R_SED}.
Given the fact that $M_{\rm BH}/M_*$ value is consistent with local relation \citep{Suh}, we may be witnessing the growing phase of both SMBH and its host galaxy in the course of the galaxy--SMBH co-evolution, as expected by the numerical simulation.

\section{Summary}
\label{Sum}

This work presents the hyper-luminous IR galaxy, eFEDSJ0828$-$0139, discovered by eROSITA.
To characterize this HyLIRG candidate, we perform the optical spectroscopy with KOOLS-IFU on the Seimei Telescope and sub-mm imaging with SCUBA-2 on JCMT.
From KOOLS-IFU observations, its spectroscopic redshift is measured to be $z_{\rm spec} = 1.622$.
We evaluate BH mass based on the single epoch method with Mg{\,\sc ii}\ line and IR luminosity based on the SED fitting.
With the caveat of the potential uncertainty of the derived physical properties discussed in sections \ref{dis_spec} and \ref{Po}, we find that IR luminosity of eFEDSJ0828$-$0139 is $L_{\rm IR} = (6.8 \pm 1.8) \times 10^{13}$ $L_{\odot}$ and Eddington ratio is $\lambda_{\rm Edd} = 3.6 \pm 0.7$, confirming an HyLIRG with SMBH being supper-Eddington accretion.
Its SFR is also high, $(1.3 \pm 0.5) \times 10^3$ $M_{\odot}$ yr$^{-1}$.
These results indicate that eFEDSJ0828$-$0139 is in a particular phase in which SMBH and its host galaxy are actively growing in the framework of galaxy-SMBH co-evolution.

Although this paper reports only one HyLIRG in the eFEDS region, several thousands of HyLIRG candidates can be selected with the aid of the eROSITA all-sky survey \citep{eRASS1}.
Spectroscopic follow-up observations with next-generation multiobject spectrographs, such as the Subaru Prime Focus Spectrograph (PFS; \cite{Takada,Greene}) provide an essential benchmark for the forthcoming systematic HyLIRG survey with eROSITA.

\begin{ack}
We acknowledge an anonymous referee for valuable suggestions that improved the paper.
We thank Drs. Masafusa Onoue and Akatoki Noboriguchi for their support in data analysis.
We also thank Dr. Fumihide Iwamuro for developing a data reduction pipeline for KOOLS-IFU. 
We are grateful to Yumiko Anraku, Yurika Matsuo, Arisa Yoshino, and the staff of the Okayama Astrophysical Observatory, a branch of the National Astronomical Observatory of Japan, for their support during our observations.
Data reduction for KOOLS-IFU was carried out on the Multi-wavelength Data Analysis System operated by the Astronomy Data Center (ADC), National Astronomical Observatory of Japan.

This work is based on data from eROSITA, the soft X-ray instrument aboard SRG, a joint Russian-German science mission supported by the Russian Space Agency (Roskosmos), in the interests of the Russian Academy of Sciences represented by its Space Research Institute (IKI), and the Deutsches Zentrum f\"ur Luft- und Raumfahrt (DLR). The SRG spacecraft was built by Lavochkin Association (NPOL) and its subcontractors, and is operated by NPOL with support from the Max Planck Institute for Extraterrestrial Physics (MPE). The development and construction of the eROSITA X-ray instrument was led by MPE, with contributions from the Dr. Karl Remeis Observatory Bamberg \& ECAP (FAU Erlangen-Nuernberg), the University of Hamburg Observatory, the Leibniz Institute for Astrophysics Potsdam (AIP), and the Institute for Astronomy and Astrophysics of the University of T\"ubingen, with the support of DLR and the Max Planck Society. The Argelander Institute for Astronomy of the University of Bonn and the Ludwig Maximilians Universit\"at Munich also participated in the science preparation for eROSITA.

The James Clerk Maxwell Telescope is operated by the East Asian Observatory on behalf of The National Astronomical Observatory of Japan; Academia Sinica Institute of Astronomy and Astrophysics; the Korea Astronomy and Space Science Institute; the National Astronomical Research Institute of Thailand; Center for Astronomical Mega-Science (as well as the National Key R\&D Program of China with No. 2017YFA0402700). Additional funding support is provided by the Science and Technology Facilities Council of the United Kingdom and participating universities and organizations in the United Kingdom and Canada.
Additional funds for the construction of SCUBA-2 were provided by the Canada Foundation for Innovation.

This work is supported by JSPS KAKENHI Grant numbers JP22H01266 and JP23K22537 (Y.T.), JP20K04027 (N.O.), and JP21K03632 (M.I.).
Z.K.G. and W.H.W. acknowledge support from the National Science and Technology Council of Taiwan (NSTC 111-2112-M-001-052-MY3).
\end{ack}

\bibliographystyle{pasj}
\bibliography{ref.bib}

\begin{thebibliography}{100}
\providecommand{\natexlab}[1]{#1}

\bibitem[{{Almeida} et~al.(2023)}]{Almeida}
{Almeida}, A. et~al. 2023, \apjs, 267, 2, 44

\bibitem[{{Assef} et~al.(2018){Assef}, {Stern}, {Noirot}, {Jun}, {Cutri}, \&
  {Eisenhardt}}]{Assef}
{Assef}, R.~J., {Stern}, D., {Noirot}, G., {Jun}, H.~D., {Cutri}, R.~M., \&
  {Eisenhardt}, P.~R.~M. 2018, \apjs, 234, 2, 23

\bibitem[{{Aufort} et~al.(2020){Aufort}, {Ciesla}, {Pudlo}, \& {Buat}}]{Aufort}
{Aufort}, G., {Ciesla}, L., {Pudlo}, P., \& {Buat}, V. 2020, \aap, 635, A136

\bibitem[{{Bailer-Jones} et~al.(2019){Bailer-Jones}, {Fouesneau}, \&
  {Andrae}}]{Bailer-Jones}
{Bailer-Jones}, C. A.~L., {Fouesneau}, M., \& {Andrae}, R. 2019, \mnras, 490,
  4, 5615

\bibitem[{{Blecha} et~al.(2018){Blecha}, {Snyder}, {Satyapal}, \&
  {Ellison}}]{Blecha}
{Blecha}, L., {Snyder}, G.~F., {Satyapal}, S., \& {Ellison}, S.~L. 2018,
  \mnras, 478, 3, 3056

\bibitem[{{Boquien} et~al.(2019){Boquien}, {Burgarella}, {Roehlly}, {Buat},
  {Ciesla}, {Corre}, {Inoue}, \& {Salas}}]{Boquien}
{Boquien}, M., {Burgarella}, D., {Roehlly}, Y., {Buat}, V., {Ciesla}, L.,
  {Corre}, D., {Inoue}, A.~K., \& {Salas}, H. 2019, \aap, 622, A103

\bibitem[{{Brunner} et~al.(2022)}]{Brunner}
{Brunner}, H. et~al. 2022, \aap, 661, A1

\bibitem[{{Bruzual} \& {Charlot}(2003)}]{Bruzual}
{Bruzual}, G. \& {Charlot}, S. 2003, \mnras, 344, 4, 1000

\bibitem[{{Buat} et~al.(2019){Buat}, {Ciesla}, {Boquien}, {Ma{\l}ek}, \&
  {Burgarella}}]{Buat}
{Buat}, V., {Ciesla}, L., {Boquien}, M., {Ma{\l}ek}, K., \& {Burgarella}, D.
  2019, \aap, 632, A79

\bibitem[{{Buendia-Rios} et~al.(2023){Buendia-Rios}, {Negrete}, {Marziani}, \&
  {Dultzin}}]{Buendia}
{Buendia-Rios}, T.~M., {Negrete}, C.~A., {Marziani}, P., \& {Dultzin}, D. 2023,
  \aap, 669, A135

\bibitem[{{Burgarella} et~al.(2005){Burgarella}, {Buat}, \&
  {Iglesias-P{\'a}ramo}}]{Burgarella}
{Burgarella}, D., {Buat}, V., \& {Iglesias-P{\'a}ramo}, J. 2005, \mnras, 360,
  4, 1413

\bibitem[{{Calderone} et~al.(2017){Calderone}, {Nicastro}, {Ghisellini},
  {Dotti}, {Sbarrato}, {Shankar}, \& {Colpi}}]{Calderone}
{Calderone}, G., {Nicastro}, L., {Ghisellini}, G., {Dotti}, M., {Sbarrato}, T.,
  {Shankar}, F., \& {Colpi}, M. 2017, \mnras, 472, 4, 4051

\bibitem[{{Calzetti} et~al.(2000){Calzetti}, {Armus}, {Bohlin}, {Kinney},
  {Koornneef}, \& {Storchi-Bergmann}}]{Calzetti}
{Calzetti}, D., {Armus}, L., {Bohlin}, R.~C., {Kinney}, A.~L., {Koornneef}, J.,
  \& {Storchi-Bergmann}, T. 2000, \apj, 533, 2, 682

\bibitem[{{Chabrier}(2003)}]{Chabrier}
{Chabrier}, G. 2003, \pasp, 115, 809, 763

\bibitem[{{Chapin} et~al.(2013){Chapin}, {Berry}, {Gibb}, {Jenness}, {Scott},
  {Tilanus}, {Economou}, \& {Holland}}]{Chapin}
{Chapin}, E.~L., {Berry}, D.~S., {Gibb}, A.~G., {Jenness}, T., {Scott}, D.,
  {Tilanus}, R. P.~J., {Economou}, F., \& {Holland}, W.~S. 2013, \mnras, 430,
  4, 2545

\bibitem[{{Ciesla} et~al.(2017){Ciesla}, {Elbaz}, \& {Fensch}}]{Ciesla}
{Ciesla}, L., {Elbaz}, D., \& {Fensch}, J. 2017, \aap, 608, A41

\bibitem[{{Dey} et~al.(2019)}]{Dey}
{Dey}, A. et~al. 2019, \aj, 157, 5, 168

\bibitem[{{Dietrich} et~al.(2002){Dietrich}, {Appenzeller}, {Vestergaard}, \&
  {Wagner}}]{Dietrich}
{Dietrich}, M., {Appenzeller}, I., {Vestergaard}, M., \& {Wagner}, S.~J. 2002,
  \apj, 564, 2, 581

\bibitem[{{Draine} et~al.(2014)}]{Draine}
{Draine}, B.~T. et~al. 2014, \apj, 780, 2, 172

\bibitem[{{Duncan}(2022)}]{Duncan}
{Duncan}, K.~J. 2022, \mnras, 512, 3, 3662

\bibitem[{{Duras} et~al.(2017)}]{Duras_WISSH}
{Duras}, F. et~al. 2017, \aap, 604, A67

\bibitem[{{Ferrarese} \& {Merritt}(2000)}]{Ferrarese}
{Ferrarese}, L. \& {Merritt}, D. 2000, \apjl, 539, 1, L9

\bibitem[{{Gao} et~al.(2021)}]{Gao}
{Gao}, F. et~al. 2021, \aap, 654, A117

\bibitem[{{Gao} et~al.(2024)}]{Zhen-Kai}
{Gao}, Z.-K. et~al. 2024, arXiv e-prints, arXiv:2405.20616

\bibitem[{{Grandi}(1982)}]{Grandi}
{Grandi}, S.~A. 1982, \apj, 255, 25

\bibitem[{{Greene} et~al.(2022){Greene}, {Bezanson}, {Ouchi}, {Silverman}, \&
  {the PFS Galaxy Evolution Working Group}}]{Greene}
{Greene}, J., {Bezanson}, R., {Ouchi}, M., {Silverman}, J., \& {the PFS Galaxy
  Evolution Working Group}. 2022, arXiv e-prints, arXiv:2206.14908

\bibitem[{{Hashiguchi} et~al.(2023)}]{Hashiguchi}
{Hashiguchi}, A. et~al. 2023, \pasj, 75, 6, 1246

\bibitem[{{Hlavacek-Larrondo} et~al.(2017)}]{Hlavacek}
{Hlavacek-Larrondo}, J. et~al. 2017, \mnras, 464, 2, 2223

\bibitem[{{Holland} et~al.(2013)}]{Holland}
{Holland}, W.~S. et~al. 2013, \mnras, 430, 4, 2513

\bibitem[{{Hoshi} et~al.(2024){Hoshi}, {Yamada}, \& {Ohta}}]{Hoshi}
{Hoshi}, A., {Yamada}, T., \& {Ohta}, K. 2024, \pasj, 76, 1, 103

\bibitem[{{Inoue}(2011)}]{Inoue}
{Inoue}, A.~K. 2011, \mnras, 415, 3, 2920

\bibitem[{{Jenness} et~al.(2008){Jenness}, {Cavanagh}, {Economou}, \&
  {Berry}}]{Jenness}
{Jenness}, T., {Cavanagh}, B., {Economou}, F., \& {Berry}, D.~S., in R.~W.
  {Argyle}, P.~S. {Bunclark}, \& J.~R. {Lewis}, eds., Astronomical Data
  Analysis Software and Systems XVII (2008), vol. 394 of \textit{Astronomical
  Society of the Pacific Conference Series}, p. 565

\bibitem[{{Kormendy} \& {Ho}(2013)}]{Kormendy}
{Kormendy}, J. \& {Ho}, L.~C. 2013, \araa, 51, 1, 511

\bibitem[{{Kurita} et~al.(2020)}]{Kurita}
{Kurita}, M. et~al. 2020, \pasj, 72, 3, 48

\bibitem[{{Lawrence} et~al.(2007)}]{Lawrence}
{Lawrence}, A. et~al. 2007, \mnras, 379, 4, 1599

\bibitem[{{Leitherer} et~al.(2002){Leitherer}, {Li}, {Calzetti}, \&
  {Heckman}}]{Leitherer}
{Leitherer}, C., {Li}, I.~H., {Calzetti}, D., \& {Heckman}, T.~M. 2002, \apjs,
  140, 2, 303

\bibitem[{{Liddle}(2004)}]{Liddle}
{Liddle}, A.~R. 2004, \mnras, 351, 3, L49

\bibitem[{{Lim} et~al.(2020)}]{Lim}
{Lim}, C.-F. et~al. 2020, \apj, 889, 2, 80

\bibitem[{{Liu} et~al.(2022)}]{Liu}
{Liu}, T. et~al. 2022, \aap, 661, A5

\bibitem[{{Lyke} et~al.(2020)}]{Lyke}
{Lyke}, B.~W. et~al. 2020, \apjs, 250, 1, 8

\bibitem[{{Magorrian} et~al.(1998)}]{Magorrian}
{Magorrian}, J. et~al. 1998, \aj, 115, 6, 2285

\bibitem[{{Marconi} \& {Hunt}(2003)}]{Marconi}
{Marconi}, A. \& {Hunt}, L.~K. 2003, \apjl, 589, 1, L21

\bibitem[{{Markwardt}(2009)}]{Markwardt}
{Markwardt}, C.~B., in D.~A. {Bohlender}, D.~{Durand}, \& P.~{Dowler}, eds.,
  Astronomical Data Analysis Software and Systems XVIII (2009), vol. 411 of
  \textit{Astronomical Society of the Pacific Conference Series}, p. 251

\bibitem[{{Martin} et~al.(2005)}]{Martin}
{Martin}, D.~C. et~al. 2005, \apjl, 619, 1, L1

\bibitem[{{Mart{\'\i}nez-Aldama} et~al.(2018){Mart{\'\i}nez-Aldama}, {del
  Olmo}, {Marziani}, {Sulentic}, {Negrete}, {Dultzin}, {D'Onofrio}, \&
  {Perea}}]{Mart}
{Mart{\'\i}nez-Aldama}, M.~L., {del Olmo}, A., {Marziani}, P., {Sulentic},
  J.~W., {Negrete}, C.~A., {Dultzin}, D., {D'Onofrio}, M., \& {Perea}, J. 2018,
  \aap, 618, A179

\bibitem[{{Marziani} \& {Sulentic}(2014)}]{Marziani}
{Marziani}, P. \& {Sulentic}, J.~W. 2014, \mnras, 442, 2, 1211

\bibitem[{{Matsubayashi} et~al.(2019)}]{Matsubayashi}
{Matsubayashi}, K. et~al. 2019, \pasj, 71, 5, 102

\bibitem[{{McConnell} \& {Ma}(2013)}]{McConnell}
{McConnell}, N.~J. \& {Ma}, C.-P. 2013, \apj, 764, 2, 184

\bibitem[{{Merloni} et~al.(2012)}]{Merloni12}
{Merloni}, A. et~al. 2012, arXiv e-prints, arXiv:1209.3114

\bibitem[{{Merloni} et~al.(2024)}]{eRASS1}
{Merloni}, A. et~al. 2024, \aap, 682, A34

\bibitem[{{Murakami} et~al.(2007)}]{Murakami}
{Murakami}, H. et~al. 2007, \pasj, 59, S369

\bibitem[{{Narayanan} et~al.(2010)}]{Narayanan}
{Narayanan}, D. et~al. 2010, \mnras, 407, 3, 1701

\bibitem[{{Noll} et~al.(2009){Noll}, {Burgarella}, {Giovannoli}, {Buat},
  {Marcillac}, \& {Mu{\~n}oz-Mateos}}]{Noll}
{Noll}, S., {Burgarella}, D., {Giovannoli}, E., {Buat}, V., {Marcillac}, D., \&
  {Mu{\~n}oz-Mateos}, J.~C. 2009, \aap, 507, 3, 1793

\bibitem[{{O'Donnell}(1994)}]{O'Donnell}
{O'Donnell}, J.~E. 1994, \apj, 422, 158

\bibitem[{{Perrotta} et~al.(2019){Perrotta}, {Hamann}, {Zakamska},
  {Alexandroff}, {Rupke}, \& {Wylezalek}}]{Perrotta}
{Perrotta}, S., {Hamann}, F., {Zakamska}, N.~L., {Alexandroff}, R.~M., {Rupke},
  D., \& {Wylezalek}, D. 2019, \mnras, 488, 3, 4126

\bibitem[{{Petrosian}(1976)}]{Petrosian}
{Petrosian}, V. 1976, \apjl, 210, L53

\bibitem[{{Predehl} et~al.(2021)}]{Predehl}
{Predehl}, P. et~al. 2021, \aap, 647, A1

\bibitem[{{Ricci} et~al.(2017)}]{Ricci}
{Ricci}, C. et~al. 2017, \apj, 835, 1, 105

\bibitem[{{Richards} et~al.(2015)}]{Richards}
{Richards}, G.~T. et~al. 2015, \apjs, 219, 2, 39

\bibitem[{{Rowan-Robinson}(2000)}]{Rowan}
{Rowan-Robinson}, M. 2000, \mnras, 316, 4, 885

\bibitem[{{Rudolf} et~al.(2016){Rudolf}, {G{\"u}nther}, {Schneider}, \&
  {Schmitt}}]{Rudolf}
{Rudolf}, N., {G{\"u}nther}, H.~M., {Schneider}, P.~C., \& {Schmitt},
  J.~H.~M.~M. 2016, \aap, 585, A113

\bibitem[{{Runnoe} et~al.(2012){Runnoe}, {Brotherton}, \& {Shang}}]{Runnoe}
{Runnoe}, J.~C., {Brotherton}, M.~S., \& {Shang}, Z. 2012, \mnras, 422, 1, 478

\bibitem[{{Salvato} et~al.(2022)}]{Salvato}
{Salvato}, M. et~al. 2022, \aap, 661, A3

\bibitem[{{Schlegel} et~al.(1998){Schlegel}, {Finkbeiner}, \& {Davis}}]{SFD}
{Schlegel}, D.~J., {Finkbeiner}, D.~P., \& {Davis}, M. 1998, \apj, 500, 2, 525

\bibitem[{{Schwarz}(1978)}]{Schwarz}
{Schwarz}, G. 1978, Annals of Statistics, 6, 2, 461

\bibitem[{{Secrest} et~al.(2015){Secrest}, {Dudik}, {Dorland}, {Zacharias},
  {Makarov}, {Fey}, {Frouard}, \& {Finch}}]{Secrest}
{Secrest}, N.~J., {Dudik}, R.~P., {Dorland}, B.~N., {Zacharias}, N., {Makarov},
  V., {Fey}, A., {Frouard}, J., \& {Finch}, C. 2015, \apjs, 221, 1, 12

\bibitem[{{Shen}(2013)}]{Shen}
{Shen}, Y. 2013, Bulletin of the Astronomical Society of India, 41, 1, 61

\bibitem[{{Shen} et~al.(2024)}]{Shen24}
{Shen}, Y. et~al. 2024, \apjs, 272, 2, 26

\bibitem[{{Stalevski} et~al.(2012){Stalevski}, {Fritz}, {Baes}, {Nakos}, \&
  {Popovi{\'c}}}]{Stalevski12}
{Stalevski}, M., {Fritz}, J., {Baes}, M., {Nakos}, T., \& {Popovi{\'c}},
  L.~{\v{C}}. 2012, \mnras, 420, 4, 2756

\bibitem[{{Stalevski} et~al.(2016){Stalevski}, {Ricci}, {Ueda}, {Lira},
  {Fritz}, \& {Baes}}]{Stalevski}
{Stalevski}, M., {Ricci}, C., {Ueda}, Y., {Lira}, P., {Fritz}, J., \& {Baes},
  M. 2016, \mnras, 458, 3, 2288

\bibitem[{{Stanley} et~al.(2015){Stanley}, {Harrison}, {Alexander}, {Swinbank},
  {Aird}, {Del Moro}, {Hickox}, \& {Mullaney}}]{Stanley}
{Stanley}, F., {Harrison}, C.~M., {Alexander}, D.~M., {Swinbank}, A.~M.,
  {Aird}, J.~A., {Del Moro}, A., {Hickox}, R.~C., \& {Mullaney}, J.~R. 2015,
  \mnras, 453, 1, 591

\bibitem[{{Suh} et~al.(2020){Suh}, {Civano}, {Trakhtenbrot}, {Shankar},
  {Hasinger}, {Sanders}, \& {Allevato}}]{Suh}
{Suh}, H., {Civano}, F., {Trakhtenbrot}, B., {Shankar}, F., {Hasinger}, G.,
  {Sanders}, D.~B., \& {Allevato}, V. 2020, \apj, 889, 1, 32

\bibitem[{{Symeonidis} \& {Page}(2018)}]{Symeonidis}
{Symeonidis}, M. \& {Page}, M.~J. 2018, \mnras, 479, 1, L91

\bibitem[{{Takada} et~al.(2014)}]{Takada}
{Takada}, M. et~al. 2014, \pasj, 66, 1, R1

\bibitem[{{Toba} et~al.(2018){Toba}, {Ueda}, {Lim}, {Wang}, {Nagao}, {Chang},
  {Saito}, \& {Kawabe}}]{Toba18}
{Toba}, Y., {Ueda}, J., {Lim}, C.-F., {Wang}, W.-H., {Nagao}, T., {Chang},
  Y.-Y., {Saito}, T., \& {Kawabe}, R. 2018, \apj, 857, 1, 31

\bibitem[{{Toba} et~al.(2015)}]{Toba15}
{Toba}, Y. et~al. 2015, \pasj, 67, 5, 86

\bibitem[{{Toba} et~al.(2020{\natexlab{a}})}]{Toba20a}
{Toba}, Y. et~al. 2020{\natexlab{a}}, \apj, 888, 1, 8

\bibitem[{{Toba} et~al.(2020{\natexlab{b}})}]{Toba20c}
{Toba}, Y. et~al. 2020{\natexlab{b}}, \apj, 899, 1, 35

\bibitem[{{Toba} et~al.(2020{\natexlab{c}})}]{Toba20b}
{Toba}, Y. et~al. 2020{\natexlab{c}}, \apj, 889, 2, 76

\bibitem[{{Toba} et~al.(2021{\natexlab{a}})}]{Toba21a}
{Toba}, Y. et~al. 2021{\natexlab{a}}, \apj, 912, 2, 91

\bibitem[{{Toba} et~al.(2021{\natexlab{b}})}]{Toba21b}
{Toba}, Y. et~al. 2021{\natexlab{b}}, \aap, 649, L11

\bibitem[{{Toba} et~al.(2022{\natexlab{a}})}]{Toba22a}
{Toba}, Y. et~al. 2022{\natexlab{a}}, \pasj, 74, 6, 1356

\bibitem[{{Toba} et~al.(2022{\natexlab{b}})}]{Toba22b}
{Toba}, Y. et~al. 2022{\natexlab{b}}, \aap, 661, A15

\bibitem[{{Toba} et~al.(2024)}]{Toba24}
{Toba}, Y. et~al. 2024, \apj, 967, 1, 65

\bibitem[{{Tody}(1986)}]{Tody}
{Tody}, D., in D.~L. {Crawford}, ed., Instrumentation in astronomy VI (1986),
  vol. 627 of \textit{Society of Photo-Optical Instrumentation Engineers (SPIE)
  Conference Series}, p. 733

\bibitem[{{V{\'e}ron-Cetty} et~al.(2004){V{\'e}ron-Cetty}, {Joly}, \&
  {V{\'e}ron}}]{Veron}
{V{\'e}ron-Cetty}, M.~P., {Joly}, M., \& {V{\'e}ron}, P. 2004, \aap, 417, 515

\bibitem[{{Vestergaard} \& {Osmer}(2009)}]{Vestergaard}
{Vestergaard}, M. \& {Osmer}, P.~S. 2009, \apj, 699, 1, 800

\bibitem[{{Vestergaard} \& {Wilkes}(2001)}]{Vestergaard01}
{Vestergaard}, M. \& {Wilkes}, B.~J. 2001, \apjs, 134, 1, 1

\bibitem[{{Vietri} et~al.(2018)}]{Vietri}
{Vietri}, G. et~al. 2018, \aap, 617, A81

\bibitem[{{Villa-V{\'e}lez} et~al.(2021){Villa-V{\'e}lez}, {Buat},
  {Theul{\'e}}, {Boquien}, \& {Burgarella}}]{Villa}
{Villa-V{\'e}lez}, J.~A., {Buat}, V., {Theul{\'e}}, P., {Boquien}, M., \&
  {Burgarella}, D. 2021, \aap, 654, A153

\bibitem[{{Wang} et~al.(2017)}]{Wang}
{Wang}, W.-H. et~al. 2017, \apj, 850, 1, 37

\bibitem[{{Wilman} et~al.(2003){Wilman}, {Fabian}, {Crawford}, \&
  {Cutri}}]{Wilman}
{Wilman}, R.~J., {Fabian}, A.~C., {Crawford}, C.~S., \& {Cutri}, R.~M. 2003,
  \mnras, 338, 1, L19

\bibitem[{{Wright} et~al.(2010)}]{Wright}
{Wright}, E.~L. et~al. 2010, \aj, 140, 6, 1868

\bibitem[{{Wu} et~al.(2023){Wu}, {Liao}, {Qi}, {Luo}, {Tang}, \& {Cao}}]{Wu23}
{Wu}, Q., {Liao}, S., {Qi}, Z., {Luo}, H., {Tang}, Z., \& {Cao}, Z. 2023,
  Research in Astronomy and Astrophysics, 23, 2, 025006

\bibitem[{{Wu} \& {Shen}(2022)}]{Wu}
{Wu}, Q. \& {Shen}, Y. 2022, \apjs, 263, 2, 42

\bibitem[{{Yang} et~al.(2020)}]{Yang}
{Yang}, G. et~al. 2020, \mnras, 491, 1, 740

\bibitem[{{Yang} et~al.(2022)}]{Yang22}
{Yang}, G. et~al. 2022, \apj, 927, 2, 192

\bibitem[{{York} et~al.(2000)}]{York}
{York}, D.~G. et~al. 2000, \aj, 120, 3, 1579

\bibitem[{{Yoshida}(2005)}]{Yoshida}
{Yoshida}, M. 2005, Journal of Korean Astronomical Society, 38, 2, 117

\bibitem[{{Yutani} et~al.(2022){Yutani}, {Toba}, {Baba}, \& {Wada}}]{Yutani}
{Yutani}, N., {Toba}, Y., {Baba}, S., \& {Wada}, K. 2022, \apj, 936, 2, 118

\end{thebibliography}

\end{document}